\documentclass{article}

\usepackage{amsmath}
\usepackage{amsfonts}
\usepackage{amssymb}
\usepackage{epsfig}

\begin{document}

\title{Field behavior of an Ising model with aperiodic interactions}

\author{Angsula Ghosh{\footnote{E-mail address: angsula@if.usp.br}},
 T. A. S. Haddad, and S. R. Salinas \\
Instituto de F\'{i}sica, Universidade de S\~{a}o Paulo, \\
Caixa Postal 66318, 05315-970, S\~ao Paulo, SP, Brazil}

\date{\today}

\maketitle

\begin{abstract}

We derive exact renormalization-group recursion relations for an Ising
model, in the presence of external fields, with ferromagnetic
nearest-neighbor interactions on Migdal-Kadanoff hierarchical lattices. We
consider layered distributions of aperiodic exchange interactions, according
to a class of two-letter substitutional sequences. For irrelevant geometric
fluctuations, the recursion relations in parameter space display a
nontrivial uniform fixed point of hyperbolic character that governs the
universal critical behavior. For relevant fluctuations, in agreement with
previous work, this fixed point becomes fully unstable, and there appears a
two-cycle attractor associated with a new critical universality class.

{PACS number(s): 05.50; 64.60; 75.10.H}
\end{abstract}

The discovery of quasi-crystals has motivated the investigation of the
effects of geometric fluctuations, associated with aperiodic interactions,
on the critical behavior of spin systems\cite{grimm}. According to the
heuristic Harris criterion\cite{harris}, it is known that fluctuations
produced by disordered interactions change the critical behavior of
ferromagnetic systems whose uniform counterparts are characterized by a
diverging specific heat at the transition. Although aperiodic interactions
come from deterministic rules, and are thus highly correlated, they also
introduce extra fluctuations that may lead to drastic changes in the
critical behavior. As proposed by Luck\cite{luck}, there is indeed a
criterion of relevance of the geometric fluctuations, which has been
confirmed by calculations for quantum spin chains with aperiodic exchange
interactions\cite{hermisson}.

Taking advantage of the structure of the Migdal-Kadanoff hierarchical
lattices, it is relatively easy to write exact recursion relations to
analyze the critical behavior of ferromagnetic Ising and Potts models, in
zero field, with layered aperiodic exchange interactions that are chosen
according to a large class of substitutional sequences\cite{s1,s2,aglae,t1}.
There is an exact analog of the Harris-Luck criterion to gauge the relevance
of the geometric fluctuations in terms of the wandering exponent of the
substitutional sequence, the number of states (of the Potts model), and some
structural features of the particular hierarchical lattice\cite{s2,aglae}.
For irrelevant fluctuations, the (universal) critical behavior is governed
by a nontrivial uniform fixed point in parameter space of the recursion
relations. For relevant fluctuations, at least in the case of two-letter
substitutional sequences, this nontrivial uniform fixed point becomes fully
unstable, and there appears a two-cycle hyperbolic attractor that gives rise
to a new class of universal critical behavior\cite{t1}.

In the present publication, we report an extension of these calculations to
analyze the field behavior of an Ising model with layered aperiodic exchange
interactions on Migdal-Kadanoff hierarchical lattices. In one dimension, we
regain results of previous work by Achiam, Lubensky, and Marshall\cite{mar}
for the ferromagnetic Ising chain with aperiodic exchange interactions
(according to the standard two-letter Fibonacci sequence). In the uniform
limit, we recover the results of Bleher and Zalys\cite{zal} for the field
behavior of the Ising model on the diamond hierarchical lattice. It should
be mentioned that spin models with uniform interactions on general
hierarchical lattices have been thoroughly investigated in a series of
papers by Kaufmann and Griffiths\cite{kauf}.

As in the work of Pinho, Haddad, and Salinas\cite{s1,s2}, we initially
consider a period-doubling two-letter sequence, 
\begin{eqnarray}
A &\rightarrow &AB,  \label{01} \\
B &\rightarrow &AA,  \label{1}
\end{eqnarray}
so that the successive application of this inflation rule on letter $A$
produces the sequence 
\begin{equation}
A\rightarrow AB\rightarrow ABAA\rightarrow ABAAABABA\rightarrow ...
\label{02}
\end{equation}
The Hamiltonian of the nearest-neighbor Ising model is written in the form 
\begin{equation}
\mathcal{H}=-\sum_{\left( i,j\right) }\left( J_{i,j}\sigma _{i}\sigma _{j}+
\frac{1}{2}h_{i,j}(\sigma _{i}+\sigma _{j})+\frac{1}{2}h_{i,j}^{\dag
}(\sigma _{i}-\sigma _{j})\right) ,  \label{ham}
\end{equation}
where $\sigma _{i}=\pm 1$ is a spin variable on site $i$ of a hierarchical
diamond structure, $\left\{ J_{ij}\right\} $ is a set of exchange
interactions, $h_{ij}$ and $h_{ij}^{\dag }$ are external fields associated
with the bonds of the structure, and the sum is over all pairs $\left(
i,j\right) $ of nearest-neighbor sites. As in the one-dimensional problem 
\cite{mar}, it is essential to take into account the staggered external
field $h^{\dag }$ in order to avoid any inconsistency in the final recursion
relations. In Fig. 1, we show three successive stages of the construction of
a diamond lattice (bond length $b=2$, and number of branches $m=2$) for the
period-doubling sequence given by Eq. (\ref{01}). As indicated in the
figure, we construct a layered distribution of both the exchange parameters, 
$J_{A}$, and $J_{B}$, and the external fields, $h_{A}$, $h_{A}^{\dag }$, 
$h_{B}$, and $h_{B}^{\dag }$, along the branches of the simple diamond
lattice. By decimating the intermediate spin variables (see Fig. 1), it is
straightforward to write the recursion relations 
\begin{equation}
x_{A}^{\prime }=\frac{
x_{A}x_{B}(1+y_{A}^{2}y_{B}^{2}z_{A}^{2}z_{B}^{2})+
y_{A}y_{B}z_{A}z_{B}(x_{B}^{2}+x_{A}^{2})}
{x_{A}x_{B}(1+y_{A}^{2}y_{B}^{2}z_{A}^{2}z_{B}^{2})
+y_{A}y_{B}z_{A}z_{B}(1+x_{B}^{2}x_{A}^{2})},  \label{rec01}
\end{equation}
\begin{equation}
x_{B}^{\prime }=\frac{
x_{A}^{2}(1+y_{A}^{4}z_{A}^{4})+2x_{A}^{2}y_{A}^{2}z_{A}^{2}}{
x_{A}^{2}(1+y_{A}^{4}z_{A}^{4})+y_{A}^{2}z_{A}^{2}(1+x_{A}^{4})},
\label{rec02}
\end{equation}
\begin{equation}
y_{A}^{\prime }=\frac{y_{A}y_{B}\left(
x_{A}x_{B}+y_{A}y_{B}z_{A}z_{B}\right) }{z_{A}z_{B}\left(
1+x_{A}x_{B}y_{A}y_{B}z_{A}z_{B}\right) },  \label{rec03}
\end{equation}
\begin{equation}
y_{B}^{\prime }=\frac{y_{A}^{2}(x_{A}^{2}+y_{A}^{2}z_{A}^{2})}{
z_{A}^{2}(1+x_{A}^{2}y_{A}^{2}z_{A}^{2})},  \label{rec04}
\end{equation}
\begin{equation}
z_{A}^{\prime }=\frac{y_{A}z_{B}(x_{B}+x_{A}y_{A}y_{B}z_{A}z_{B})}{
y_{B}z_{A}(x_{A}+x_{B}y_{A}y_{B}z_{A}z_{B})},  \label{rec05}
\end{equation}
and 
\begin{equation}
z_{B}^{\prime }=1,  \label{rec06}
\end{equation}
where $x_{A,B}=\exp \left( -2\beta J_{A,B}\right) $, $y_{A,B}=\exp \left(
-\beta h_{A,B}\right) $, $z_{A,B}=\exp \left( -\beta h_{A,B}^{\dag }\right)$
, and $\beta $ is the inverse of the temperature.

In the uniform case, $J_{A}=J_{B}=J$, $h_{A}=h_{B}=h$, and $h_{A}^{\dag
}=h_{B}^{\dag }=h^{\dag }$, Eqs. (\ref{rec01})-(\ref{rec06}) reduce to the
recursion relations 
\begin{equation}
x^{\prime }=\frac{x^{2}(1+y^{4}z^{4})+2x^{2}y^{2}z^{2}}
{x^{2}(1+y^{4}z^{4})+y^{2}z^{2}(1+x^{4})},  \label{recb01}
\end{equation}
\begin{equation}
y^{\prime }=\frac{y^{2}}{z^{2}}\frac{x^{2}+y^{2}z^{2}}{1+x^{2}y^{2}z^{2}},
\label{recb02}
\end{equation}
and 
\begin{equation}
z^{\prime }=1,  \label{recb03}
\end{equation}
which can be compared with the expressions obtained by Bleher and Zalys\cite
{zal} for the simple Ising model in a (uniform) field. For the aperiodic
model, but in the absence of the external fields, we recover the equations
obtained by Pinho et al.\cite{s1,s2}. Due to the presence of a staggered
field, it is important to see that the form of the recursion relations
depends on the way we label the bonds on the diamond lattice. To be
consistent with the calculations of Bleher and Zalys, we assume a positive
staggered field at the intermediate spins ($\sigma _{i}$) of the polygon,
and an opposite field at the nodes ($\sigma _{j}$). With a change of order,
there is no automatic cancellation of the staggered field associated with
the $B$ bonds.

The nontrivial uniform fixed point of the recursion relations (\ref{rec01})-
(\ref{rec06}) is given by $y_{A}=y_{B}=z_{A}=z_{B}=1$, corresponding to an
Ising model in zero field ($h_{A}=h_{B}=h_{A}^{\dag }=h_{B}^{\dag }=0$), and 
\begin{equation}
x_{A}^{\ast }=x_{B}^{\ast }=x=0.295598...,
\end{equation}
which comes from the cubic equation $x^{3}+x^{2}+3x-1=0$, and is the usual
result for the uniform model. The trivial fixed points are located at 
$y_{A}=y_{B}=z_{A}=z_{B}=1$, with $x_{A}^{\ast }=x_{B}^{\ast }=0$ or $1$. The
linearization of the recursion relations about the nontrivial uniform fixed
point, yields the block-diagonal form 
\begin{equation}
\left( 
\begin{array}{c}
\Delta x_{A}^{\prime } \\ 
\Delta x_{B}^{\prime } \\ 
\Delta y_{A}^{\prime } \\ 
\Delta y_{B}^{\prime } \\ 
\Delta z_{A}^{\prime } \\ 
\Delta z_{B}^{\prime }
\end{array}
\right) =\left( 
\begin{array}{cccccc}
C & C & 0 & 0 & 0 & 0 \\ 
2C & 0 & 0 & 0 & 0 & 0 \\ 
0 & 0 & 1+C & 1+C & -1+C & -1+C \\ 
0 & 0 & 2+2C & 0 & -2+2C & 0 \\ 
0 & 0 & 1 & -1 & -1 & 1 \\ 
0 & 0 & 0 & 0 & 0 & 0
\end{array}
\right) \left( 
\begin{array}{c}
\Delta x_{A} \\ 
\Delta x_{B} \\ 
\Delta y_{A} \\ 
\Delta y_{B} \\ 
\Delta z_{A} \\ 
\Delta z_{B}
\end{array}
\right) ,
\end{equation}
where $C=(1-x^{2})/(1+x^{2})=0.839286...$. The thermal eigenvalues, given by 
\begin{equation}
\Lambda _{1}=2C=1.67857...,\quad \text{and }\Lambda _{2}=-C=-0.839286...,
\end{equation}
are identical to the findings of Pinho, Haddad, and Salinas\cite{s1}, in
zero external field. The magnetic eigenvalues are given by 
\begin{equation}
\Lambda _{3}=2\left( 1+C\right) =3.678572...,\text{ \qquad }\Lambda _{4}=-2,
\text{ }
\end{equation}
\begin{equation}
\Lambda _{5}=-C=-0.839286...,\text{ \qquad and }\Lambda _{6}=0.
\end{equation}
For the Ising model on the diamond lattice with a uniform distribution of
exchange interactions (as in the work of Bleher and Zalys), the
linearization about the nontrivial fixed point of Eqs. (\ref{recb01})-(\ref
{recb02}) leads to the set of three uniform eigenvalues, $\Lambda
_{1}=2C=1.67857...$, $\Lambda _{3}=2\left( 1+C\right) =3.678572...$, and $
\Lambda _{6}=0$. Therefore, as in the work of Pinho, Haddad, and Salinas,
and in agreement with the analog of the Harris-Luck criterion, we see that
the geometric fluctuations associated with the period-doubling sequence are
unable to change the ferromagnetic critical behavior of the Ising model on
the diamond lattice. The absolute value of the smaller thermal eigenvalue is
less than $1$. The absolute values of the eigenvalues associated with the
uniform field are both larger than one. The staggered field is irrelevant as
in the uniform case. Also, it should be noted that the same kind of behavior
still holds for a more general diamond-type lattice, with $b=2$ bonds and $
m>2$ branches.

Due to the peculiar structure of the diamond hierarchical lattice, there is
a change in the magnetic eigenvalues if we make another choice for the sign
of the staggered field. This effect is already present in the model with
uniformly distributed exchange interactions. In fact, if we linearize the
new recursion relations of the uniform model, in direct and staggered
applied fields, we obtain the same values for $\Lambda _{1}$ and $\Lambda
_{3}$, but $\Lambda _{6}=-2$, instead of $\Lambda _{6}=0$. For the aperiodic
model, we obtain the new values $\Lambda _{4}=-1.78151...$, $\Lambda
_{5}=-0.942221...$, and $\Lambda _{6}=-2$. Hence, we see that we cannot
attribute this change of magnetic eigenvalues to the introduction of
aperiodic exchange interactions. It is evident that aperiodicity is not the
cause for these changes.

To give an example of relevant geometric fluctuations, we now consider the
period-three two-letter sequence, 
\begin{equation}
A\rightarrow ABB,
\end{equation}
\begin{equation}
B\rightarrow AAA.
\end{equation}
The nearest-neighbor Ising model is defined on a Migdal-Kadanoff lattice
with $b=3$ bonds and $m=2$ branches. Although somewhat tedious, it is
straightforward to write the recursion relations 
\begin{equation}
x_{A}^{\prime }=\frac{F_{1}\left( x_{A},x_{B},y_{A},y_{B},z_{A,}z_{B}\right)
F_{2}\left( x_{A},x_{B},y_{A},y_{B},z_{A,}z_{B}\right) }{F_{3}\left(
x_{A},x_{B},y_{A},y_{B},z_{A,}z_{B}\right)
F_{4}(x_{A},x_{B},y_{A},y_{B},z_{A,}z_{B})},  \label{recc01}
\end{equation}
\begin{equation}
x_{B}^{\prime }=\frac{F_{1}\left( x_{A},x_{A},y_{A},y_{A},z_{A,}z_{A}\right)
F_{2}\left( x_{A},x_{A},y_{A},y_{A},z_{A,}z_{A}\right) }{F_{3}\left(
x_{A},x_{A},y_{A},y_{A},z_{A,}z_{A}\right) F_{4}\left(
x_{A},x_{A},y_{A},y_{A},z_{A,}z_{A}\right) },  \label{recc02}
\end{equation}
\begin{equation}
y_{A}^{\prime }=\frac{y_{A}y_{B}z_{A}F_{4}\left(
x_{A},x_{B},y_{A},y_{B},z_{A,}z_{B}\right) }{z_{B}F_{3}\left(
x_{A},x_{B},y_{A},y_{B},z_{A,}z_{B}\right) },  \label{recc03}
\end{equation}
\begin{equation}
y_{B}^{\prime }=\frac{y_{A}^{2}F_{4}\left(
x_{A},x_{A},y_{A},y_{A},z_{A,}z_{A}\right) }{F_{3}\left(
x_{A},x_{A},y_{A},y_{A},z_{A,}z_{A}\right) },  \label{recc04}
\end{equation}
\begin{equation}
z_{A}^{\prime }=\frac{y_{A}z_{A}z_{B}F_{1}\left(
x_{A},x_{B},y_{A},y_{B},z_{A,}z_{B}\right) }{y_{B}F_{2}\left(
x_{A},x_{B},y_{A},y_{B},z_{A,}z_{B}\right) },  \label{recc05}
\end{equation}
\begin{equation}
z_{B}^{\prime }=\frac{z_{A}^{2}F_{1}\left(
x_{A},x_{A},y_{A},y_{A},z_{A,}z_{A}\right) }{F_{2}\left(
x_{A},x_{A},y_{A},y_{A},z_{A,}z_{A}\right) },  \label{recc06}
\end{equation}
where 
\begin{equation}
F_{1}\left( x_{A},x_{B},y_{A},y_{B},z_{A,}z_{B}\right)
=x_{A}z_{A}z_{B}+x_{A}x_{B}^{2}y_{B}^{2}z_{A}z_{B}^{3}+x_{B}
y_{A}y_{B}+x_{B}y_{A}y_{B}^{3}z_{B}^{2},
\end{equation}
\begin{equation}
F_{2}\left( x_{A},x_{B},y_{A},y_{B},z_{A,}z_{B}\right)
=x_{B}z_{A}z_{B}+x_{A}x_{B}^{2}y_{A}y_{B}+x_{B}y_{B}^{2}
z_{A}z_{B}^{3}+x_{A}y_{A}y_{B}^{3}z_{B}^{2},
\end{equation}
\begin{equation}
F_{3}\left( x_{A},x_{B},y_{A},y_{B},z_{A,}z_{B}\right)
=z_{A}z_{B}+x_{A}x_{B}y_{A}y_{B}+x_{A}x_{B}y_{A}y_{B}^
{3}z_{B}^{2}+x_{B}^{2}y_{B}^{2}z_{A}z_{B}^{3},
\end{equation}
and 
\begin{equation}
F_{4}\left( x_{A},x_{B},y_{A},y_{B},z_{A,}z_{B}\right)
=x_{A}x_{B}z_{A}z_{B}+x_{A}x_{B}y_{B}^{2}z_{A}z_{B}
^{3}+x_{B}^{2}y_{A}y_{B}+y_{A}y_{B}^{3}z_{B}^{2}.
\end{equation}

Besides the trivial fixed points, there is a nontrivial (uniform) fixed
point, given by $y_{A}=y_{B}=z_{A}=z_{B}=1$ (zero direct and staggered
fields), and 
\begin{equation}
x_{A}^{\ast }=x_{B}^{\ast }=x=0.119726...,  \label{ntf1}
\end{equation}
which comes from the equation 
\begin{equation}
x=\left( \frac{1+3x^{2}}{3+x^{2}}\right) ^{2},
\end{equation}
and is identical to the fixed point associated with the uniform model. We
now linearize the recursion relations about this nontrivial fixed point.
Again, we have a block-diagonal form involving separate thermal and magnetic
contributions, 
\begin{equation}
\left( 
\begin{array}{c}
\Delta x_{A}^{\prime } \\ 
\Delta x_{B}^{\prime }
\end{array}
\right) =\left( 
\begin{array}{cc}
D & 2D \\ 
3D & 0
\end{array}
\right) \left( 
\begin{array}{c}
\Delta x_{A} \\ 
\Delta x_{B}
\end{array}
\right) ,
\end{equation}
where 
\begin{equation}
D=\frac{2\left( 1-x^{2}\right) ^{2}}{\left( 3+x^{2}\right) \left(
1+3x^{2}\right) }=0.618033...,
\end{equation}
and 
\begin{equation}
\left( 
\begin{array}{c}
\Delta y_{A}^{\prime } \\ 
\Delta y_{B}^{\prime } \\ 
\Delta z_{A}^{\prime } \\ 
\Delta z_{B}^{\prime }
\end{array}
\right) =\left( 
\begin{array}{cccc}
2\left( 1+x^{2}\right) E & 4E & 4x^{2}E & -4x^{2}E \\ 
2\left( 3+x^{2}\right) E & 0 & 0 & 0 \\ 
4F & -4F & 2\left( 1+x^{2}\right) F & 4x^{2}F \\ 
0 & 0 & 2\left( 1+3x^{2}\right) F & 0
\end{array}
\right) \left( 
\begin{array}{c}
\Delta y_{A} \\ 
\Delta y_{B} \\ 
\Delta z_{A} \\ 
\Delta z_{B}
\end{array}
\right) ,
\end{equation}
where 
\begin{equation}
E=\frac{1}{1+3x^{2}}\qquad \text{and\qquad }F=\frac{1}{3+x^{2}}.
\end{equation}
We then have the thermal eigenvalues, 
\begin{equation}
\Lambda _{1}=3D=1.854101...,\quad \text{and }\Lambda _{2}=-2D=-1.236067...,
\end{equation}
which are identical to the findings of Pinho, Haddad, and Salinas\cite{s1},
in zero external field, and the magnetic eigenvalues, 
\begin{equation}
\Lambda _{3}=\frac{2E}{F}=5.780107...,\text{ \quad }\Lambda _{4}=-4\left(
E+x^{2}F\right) =-3.854101...,
\end{equation}
\begin{equation}
\Lambda _{5}=0,\quad \text{and\quad }\Lambda _{6}=\frac{2F}{E}=0.692028....
\end{equation}
For the Ising model on the diamond lattice with a uniform distribution of
exchange interactions (as in the work of Bleher and Zalys), the
linearization about the nontrivial fixed point of Eqs. (\ref{recc01})-(\ref
{recc06}) leads to the set of three uniform eigenvalues, $\Lambda _{1}=3D>1$
, $\Lambda _{3}=2E/F>1$, and $0<\Lambda _{6}=2F/E<1$ (the external staggered
field is irrelevant). As in the work of Pinho, Haddad, and Salinas\cite{s1},
since $\Lambda _{1}>1$ and $\left| \Lambda _{2}\right| >1$, the uniform
fixed point is unstable along both thermal directions, and cannot be reached
from physically acceptable initial conditions. This type of behavior, in
full agreement with the exact analog of the Harris-Luck criterion, still
holds for more general lattices with $b=3$ but $m>2$.

In the more recent analyses of Haddad, Pinho, and Salinas\cite{t1}, in
addition to the (unstable) uniform fixed point, there is a two-cycle
attractor that governs the critical behavior. Recursion relations (\ref
{recc01})-(\ref{recc06}) are also associated with a two-cycle attractor,
located at $y_{A}=y_{B}=z_{A}=z_{B}=1$, and 
\begin{equation}
x_{A}^{\ast }=0.155117...,\quad x_{B}^{\ast }=0.00740154...,
\end{equation}
and 
\begin{equation}
x_{A}^{\ast }=0.0287405...,\quad x_{B}^{\ast }=0.191409....
\end{equation}
The linearization of the second iterates of the recursion relations about
this two-cycle attractor leads to the thermal eigenvalues, 
\begin{equation}
\Lambda _{1}=3.255710...,\text{ and }\Lambda _{2}=0.311746...,
\end{equation}
where $\left| \Lambda _{2}\right| <1$ indicates that the attractor can be
reached from physically acceptable initial conditions, and the magnetic
eigenvalues 
\begin{equation}
\Lambda _{3}=34.384598...,\qquad \Lambda _{4}=15.008653...,
\end{equation}
and 
\begin{equation}
\Lambda _{5}=0.267660...,\qquad \text{and \quad }\Lambda _{6}=0,
\end{equation}
which indicates the irrelevance of the applied staggered field. These
results are consistent with the recent work of Haddad, Pinho, and Salinas 
\cite{t1}. The nontrivial diagonal fixed point cannot be reached. In zero
field, a new universal critical behavior is characterized by the eigenvalues
about the cycle-two attractor.

In conclusion, we have investigated a ferromagnetic Ising model with
aperiodic exchange interactions, in the presence of direct and staggered
external fields, on a class of Migdal-Kadanoff hierarchical structures. The
geometric fluctuations introduced by the aperiodic interactions may change
the character of the fixed point associated with the critical behavior. For
relevant geometric fluctuations, the critical behavior belongs to a new
universality class associated with a two-cycle attractor. The inclusion of a
magnetic field does confirm these findings.

The authors thank Funda\c c\~ao de Amparo \`a Pesquisa do Estado de S\~ao
Paulo for financial support.

\newpage

\begin{figure}
\begin{center}
\epsfbox{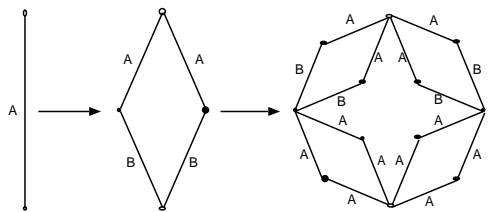}
\end{center}
\caption{Some generations of the simple diamond hierarchical lattice ($b=2$
bonds and $m=2$ branches). The layered exchange interactions are chosen
according to the period-doubling two-letter substitution sequence, 
$A\rightarrow AB$, $B\rightarrow AA$.}
\end{figure}

\end{document}